\documentclass{ws-ijmpa}

\begin{document}

\markboth{H. Kawamura et al.}
{$Q_T$ resummation in Transversely Polarized Drell-Yan process}

\catchline{}{}{}{}{}

\title{$Q_T$ RESUMMATION IN TRANSVERSELY POLARIZED DRELL-YAN PROCESS
\footnote{A talk presented by H.Kawamura
% at International Conference 
%on QCD and Hadron Physics, June.16-20,2005, Beijing
}}

\author{HIROYUKI KAWAMURA}

\address{Radiation Laboratory, RIKEN\\ 
2-1 Hirosawa, Wako, Saitama 351-0198, JAPAN\\ 
kawamura@rarfaxp.riken.jp}

\author{JIRO KODAIRA and HIROTAKA SHIMIZU}

\address{Theory Division, High Energy Accelerator Research Organization (KEK)\\
1-1 OHO, Tsukuba 305-0801, JAPAN}

\author{KAZUHIRO TANAKA}

\address{Department of Physics, Juntendo University, Inba-gun, 
Chiba 270-1695, JAPAN}

%%%%%%%%%%%%%%%%%%%%%%%%%%%%%%%%%%%%%%%%%%%%%%%%%%%%%%%%%%%%
% You may repeat \author \address as often as necessary    %
%%%%%%%%%%%%%%%%%%%%%%%%%%%%%%%%%%%%%%%%%%%%%%%%%%%%%%%%%%%%

\maketitle

\begin{abstract}
We calculate QCD corrections to transversely polarized Drell-Yan 
process at a measured $Q_T$ of the produced lepton pair
in the dimensional regularization scheme.
The $Q_T$ distribution is discussed resumming soft gluon 
effects relevant for small $Q_T$.
\end{abstract}

\keywords{$Q_T$ resummation, Transversity, Drell-Yan process}

%%%%%%%%%%%%%%%%%%%%%%%%%%%%%%%%%%%%%%%%%%%%%%%%%%%%%%%%%%%%
% The main text of your paper	begins here                %
%%%%%%%%%%%%%%%%%%%%%%%%%%%%%%%%%%%%%%%%%%%%%%%%%%%%%%%%%%%%
\vspace{0.4cm}
\noindent
Hard processes with polarized nucleon beams enable us to study
spin-dependent dynamics of QCD and the spin structure of nucleon. 
The helicity distribution $\Delta q(x)$ of quarks within nucleon 
has been measured in polarized DIS experiments, and $\Delta G(x)$ of gluons
has also been estimated from the scaling violations of them. 
On the other hand, the transversity distribution $\delta q(x)$, 
i.e. the distribution of transversely polarized quarks inside
transversely polarized nucleon, can not be measured in inclusive DIS 
due to its chiral-odd nature,\cite{RS:79} 
and remains as the last unknown distribution at the leading twist. 
Transversely polarized Drell-Yan (tDY) process is one of the processes 
where the transversity distribution can be measured,
and has been undertaken at RHIC-Spin experiment.
 
We compute the 1-loop QCD corrections to tDY at a measured $Q_T$ 
and azimuthal angle $\phi$ of the produced lepton in the dimensional 
regularization scheme.
For this purpose, the phase space integration in $D$-dimension, 
separating out the relevant transverse degrees of freedom, 
is required to extract the $\propto \! \cos(2\phi)$ part of the cross
section characteristic of the spin asymmetry of tDY.\cite{RS:79}
The calculation is rather cumbersome compared with the corresponding 
calculation in unpolarized and longitudinally polarized cases,
and has not been performed so far.
We obtain the NLO ${\cal O}(\alpha_s)$ corrections to the tDY
cross section in the $\overline{\rm MS}$ scheme.
We also include soft gluon effects by all-order resummation of 
logarithmically enhanced contributions at small $Q_T$ 
(``edge regions of the phase space'')
up to next-to-leading logarithmic (NLL) accuracy, and 
obtain the first complete result of the $Q_T$ distribution for all 
regions of $Q_T$ at NLL level.
 
We first consider the NLO ${\cal O}(\alpha_s)$ corrections to tDY:
$h_1(P_1,s_1)+h_2(P_2,s_2)\rightarrow l(k_1)+\bar{l}(k_2)+X$,
where $h_1,h_2$ denote nucleons with momentum $P_1,P_2$ and transverse 
spin $s_1,s_2$, and $Q=k_1+k_2$ is the 4-momentum of DY pair.  
The spin dependent cross section
$\Delta_T d \sigma \equiv (d \sigma (s_1 , s_2) - d \sigma (s_1 , - s_2))/2$
is given as a convolution
\begin{equation}
\Delta_T d\sigma = \int d x_1 d x_2\,  
    \delta H (x_1 \,,\,x_2 ; \mu_F)\, 
             \Delta_T d \hat{\sigma} (s_1\,,\,s_2 ; \mu_F), 
\end{equation}
where $\mu_F$ is the factorization scale, and
\begin{equation}
\delta H (x_1 \,,\,x_2 ; \mu_F)\, = \sum_i e_i^2 
[\delta q_i(x_1 ; \mu_F)\delta \bar{q}_i(x_2 ; \mu_F)
+\delta \bar{q}_i(x_1 ; \mu_F)\delta q_i(x_2 ; \mu_F)]
\end{equation}
is the product of transversity distributions of the two nucleons,
and $\Delta_T d \hat{\sigma}$ is the corresponding partonic 
cross section. 
Note that, at the leading twist level, the gluon does not contribute 
to the transversely polarized process due to its chiral odd nature. 
We compute the one-loop corrections to $\Delta_T d \hat{\sigma}$,
which involve the virtual gluon corrections and the real gluon emission 
contributions, e.g.,
$q (p_1 , s_1) + \bar{q} (p_2 , s_2) 
              \to l (k_1) + \bar{l} (k_2) + g$,
with $p_i = x_i P_i$.
We regularize the infrared divergence  
in $D=4 - 2 \epsilon$ dimension, and employ naive anticommuting $\gamma_5$
which is a usual prescription in the transverse spin channel.\cite{WV:98}
In the $\overline{\rm MS}$ scheme, we eventually get,\cite{KKST,KKST2} to NLO accuracy, 
\begin{eqnarray}
\frac{\Delta_T d \sigma}{d Q^2 d Q_T^2 d y d \phi}
= N\, \cos{(2 \phi )} 
 \left[ X\, (Q_T^2 \,,\, Q^2 \,,\, y) 
+ Y\, (Q_T^2 \,,\, Q^2 \,,\, y) \right],
\label{cross section}
\end{eqnarray}
where $N = \alpha^2 / (3\, N_c\, S\, Q^2)$ with $S=(P_1 +P_2 )^2$,
$y$ is the rapidity of virtual photon, and $\phi$ is the azimuthal 
angle of one of the leptons with respect to the initial spin axis.
For later convenience, we have decomposed the cross section into 
the two parts: the function $X$ contains all terms that are singular 
as $Q_T \rightarrow 0$, while $Y$ is of ${\cal O}(\alpha_s)$ and  
finite at $Q_T=0$.
Writing $X = X^{(0)} + X^{(1)}$ as the sum of the LO and 
NLO contributions,  we have\cite{KKST,KKST2}
$X^{(0)} = \delta H (x_1^0\,,\,x_2^0\,;\, \mu_F )\ \delta (Q_T^2)$,
and
\begin{eqnarray}
X^{(1)} &=& \frac{\alpha_s}{2 \pi} C_F\ 
       \Biggl\{ \delta H (x_1^0\,,\,x_2^0\,;\, \mu_F ) 
    \left[\, 2\, \left( \frac{\ln Q^2 / Q_T^2}{Q_T^2} \right)_+ 
              - \frac{3}{(Q_T^2)_+}
    + \left(\, - 8 + \pi^2 \right) \delta (Q_T^2) \right] 
\nonumber\\
 \!\!\!\!\!\!\!\!\!\!\!\!\!\!\!\!\!+&& \!\!\!\!\!\!\!\!\left(  \frac{1}{(Q_T^2)_+} 
        + \delta (Q_T^2) \ln \frac{Q^2}{\mu_F^2} \right)\!\!\!
       \left[ \int^1_{x_1^0} \frac{d z}{z}
\delta P_{qq}^{(0)} (z)\
        \delta H 
     \left( \frac{x_1^0}{z}, x_2^0 ;\ \mu_F \right)
        +  ( x_1^0 \leftrightarrow x_2^0 ) \right] \Biggr\} ,
\label{eq:x}
\end{eqnarray}
where $x_1^0 = \sqrt{\tau}\ e^y , x_2^0 =\sqrt{\tau}\ e^{-y}$ 
are the relevant scaling variables with $\tau =Q^2/S$,
and $\delta P_{qq}^{(0)} (z) = 2 z/(1 - z)_+ 
            + (3/2)\, \delta (1 - z)$
is the LO transverse splitting function.\cite{AM:90}
In (\ref{eq:x}), the terms involving $\delta(Q_T^2 )$ come from the 
virtual gluon corrections, while the other terms represent the recoil 
effects due to the real gluon emissions.
For the analytic expression of $Y$, see Ref.\cite{KKST2}.
Eq.~(\ref{cross section}) gives the first NLO result 
in the $\overline{\rm MS}$ scheme.
We note that there has been a similar NLO calculation of tDY
cross section in massive gluon scheme.\cite{VW:93}
We also note that, integrating (\ref{cross section}) over $Q_T$,
our result coincides with the corresponding $Q_T$-integrated 
cross sections obtained in previous works employing massive 
gluon scheme\cite{VW:93} and dimensional reduction scheme,\cite{CKM:94} 
via the scheme transformation relation.\cite{WV:98}

The cross section (\ref{cross section}) becomes very large when 
$Q_T \ll Q$, due to the terms behaving 
$\sim \alpha_s \ln(Q^2/Q_T^2 )/Q_T^2$ and $\sim \alpha_s /Q_T^2$
in the singular part $X$.
It is well-known that, in unpolarized and longitudinally polarized DY,
large ``recoil logs'' of similar nature appear in each order of 
perturbation theory as 
$\alpha_s^n \ln^{2n-1}(Q^2/Q_T^2 )/Q_T^2$, 
$\alpha_s^n \ln^{2n-2}(Q^2/Q_T^2 )/Q_T^2$, and so on,
corresponding to LL, NLL, and higher contributions, respectively,
and that the resummation of those ``double logarithms'' to 
all orders is necessary to obtain a well-defined, finite 
prediction of the cross section.\cite{CSS:85} 
Because the LL and NLL contributions are universal,\cite{DG:00}
we can work out the all-order resummation of the corresponding 
logarithmically enhanced contributions
in (\ref{cross section}) up to the NLL accuracy, 
based on the general formulation\cite{CSS:85} of the $Q_T$ resummation.
This can be conveniently carried out in the impact parameter $b$ space,
conjugate to the $Q_T$ space.
As a result, the singular part $X$ of (\ref{cross section}) is modified 
into the corresponding resummed part, which is expressed as the Fourier 
transform,
\begin{eqnarray}
X \rightarrow&&
\sum_{i}e_i^2 \int_0^{\infty} d b \frac{b}{2}
J_0 (b Q_T)
     e^{\, S (b , Q)} 
( C_{qq} \otimes \delta q_i )
           \left( x_1^0 , \frac{b_0^2}{b^2} \right)      
 ( C_{\bar{q} \bar{q}} \otimes \delta\bar{q}_i )
           \left( x_2^0 , \frac{b_0^2}{b^2} \right) 
\nonumber\\
&&+ ( x_1^0 \leftrightarrow x_2^0 ) .
\label{resum}
\end{eqnarray}
Here $b_0 = 2e^{-\gamma_E}$, 
and the large logarithmic corrections are resummed into the 
Sudakov factor $e^{S (b , Q)}$ with
$S(b,Q)=-\int_{b^2_0/b^2}^{Q^2}(d\kappa^2/\kappa^2)
\{ (\ln{\frac{Q^2}{\kappa^2}} )A_q(\alpha_s(\kappa))+
B_q(\alpha_s(\kappa)) \}$.
The functions $A_q$, $B_q$ as well as the
coefficient functions $C_{qq}, C_{\bar{q}\bar{q}}$
are calculable in perturbation theory, and
at the present accuracy of NLL, we get:\cite{KKST,KKST2}
$A_q(\alpha_s ) =(\alpha_s /\pi) C_F +(\alpha_s /2\pi)^2 
2C_F \{(67/18-\pi^2/6)C_G-5N_f /9 \}$,
$B_q (\alpha_s )=-3C_F(\alpha_s /2\pi)$,
$C_{qq}(z, \alpha_s )=C_{\bar{q} \bar{q}}(z, \alpha_s )
=\delta(1-z)\{1+(\alpha_s /4\pi)C_F(\pi^2-8)\}$.
We have utilized a relation\cite{KT:82} between $A_q$ and 
the DGLAP kernels in order to obtain the two-loop term of $A_q$.
The other contributions have been determined so that
the expansion of the above formula (\ref{resum}) in powers
of $\alpha_s (\mu_F )$ reproduces $X$ of (\ref{cross section}), 
(\ref{eq:x}) to the NLO accuracy.
Eq.~(\ref{cross section}) with (\ref{resum}) presents the first 
result of the NLL $Q_T$ resummation formula for tDY. 
The NLO parton distributions in the $\overline{\rm MS}$ scheme
have to be used.

One more step is necessary to make the QCD prediction 
of tDY. Similarly to other all-order resuumation formula, 
our result (\ref{resum}) is suffered from the IR renormalons 
due to the Landau pole at $b= (b_0 /Q)e^{(1/2\beta_0 \alpha_s (Q))}$
in the Sudakov factor, and
it is necessary to specify a prescription to avoid 
this singularity. Here we deform the integration contour in 
(\ref{resum}) in the complex $b$ space, following the method 
introduced in the joint resummation.\cite{LKSV:01}  
Obviously prescription to define the $b$ integration
is not unique reflecting IR renormalon ambiguity,
e.g., 
``$b_{*}$ prescription'' to ``freeze'' effectively the $b$ integration  
along the real axis is frequently used.\cite{CSS:85}
The renormalon ambiguity should be eventually compensated in 
the physical quantity by the power corrections
$\sim (b \Lambda_{\rm  QCD})^n$ ($n=2,3, \ldots$) 
due to non-perturbative effects. 
Correspondingly, we make the replacement 
$e^{S (b , Q)}\rightarrow e^{S (b , Q)}F^{NP}(b)$ in (\ref{resum}) 
with the ``minimal'' ansatz for non-perturbative effects,
\cite{CSS:85,LKSV:01} 
$F^{NP}(b)=\exp(-g b^2)$
with a non-perturbative parameter $g$.
Fig.1 shows the $Q_T$ distribution of tDY at $\sqrt{S}=100$ GeV, 
$Q=10$ GeV, $y=\phi=0$, and with a model for the transversity 
$\delta q(x)$ that saturates the Soffer's inequality 
at a low scale.\cite{MSSV:98} 
Solid line shows the NLO result using (\ref{cross section}), and 
the dashed and dot-dashed lines show the NLL result using 
(\ref{cross section}), (\ref{resum}), $F^{NP}(b)=\exp(-g b^2)$, 
with $g=0.5$ GeV$^2$ and $g=0$, respectively.
 
\begin{figure}
\centerline{\psfig{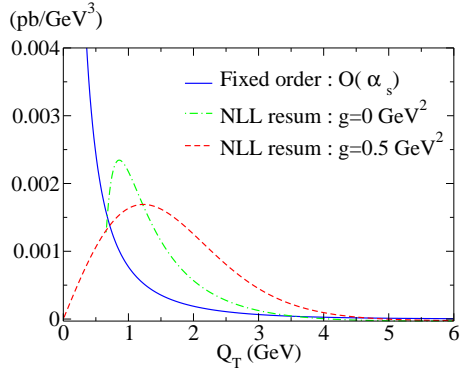}}
\vspace*{8pt}
\caption{$Q_T$ distribution $\Delta_T d \sigma/d Q^2 d Q_T d y d \phi$
at $\sqrt{S}=100$ GeV, $Q=10$ GeV, $y=\phi=0$. }
\end{figure}

%%%%%%%%%%%%%%%%%%%%%%%%%%%%%%%%%%%%%%%%%%%%%%%%%%%%%%%%%%%%
% Doing Acknowledgement			                   %
%%%%%%%%%%%%%%%%%%%%%%%%%%%%%%%%%%%%%%%%%%%%%%%%%%%%%%%%%%%%

\vspace{-0.1cm}

\section*{Acknowledgements}
We would like to thank W. Vogelsang for valuable discussions. 
The work of J.K. was supported by the Grant-in-Aid for 
Scientific Research No. C-16540255. The work of K.T. was 
supported by the Grant-in-Aid for Scientific Research No. C-16540266. 

%%%%%%%%%%%%%%%%%%%%%%%%%%%%%%%%%%%%%%%%%%%%%%%%%%%%%%%%%%%%
% Doing references:	                   	           %
%%%%%%%%%%%%%%%%%%%%%%%%%%%%%%%%%%%%%%%%%%%%%%%%%%%%%%%%%%%%

\end{document}